\documentclass[12pt]{iopart}
\usepackage{graphicx}% Include figure files
\usepackage{bm}% bold math
\usepackage{iopams}
\usepackage{bbm}
\usepackage{mathrsfs} % cool fonts: try $\mathscr P$
\usepackage[all]{xy}

\usepackage[usenames]{color}

\newcommand{\col}[1]{#1} %operator

\newcommand{\image}[4]{
\begin{figure}
   \begin{center}
      \includegraphics[angle = 0, width = #2]{#1}
   \end{center}
\caption{\label{#4}#3}
\end{figure}
}

\newcommand{\xnode}[1]{*+<0.4pc>[F]{#1}}

\newcommand{\xminid}[1]{\xymatrix@C=5mm{#1}}
\newcommand{\Par}{\ar@{-}[r]}
\newcommand{\Pau}{\ar@{-}[u]}
\newcommand{\Pad}{\ar@{-}[d]}
\newcommand{\Pal}{\ar@{-}[l]}
\newcommand{\bra}[1]{\langle #1|}
\newcommand{\ket}[1]{| #1 \rangle}
\newcommand{\braket}[2]{\langle #1|#2 \rangle}
\newcommand{\ketbra}[2]{| #1 \rangle\langle#2 |}
\newcommand{\expi}[1]{ {\rm e}^{{\rm i} #1}}
\newcommand{\expim}[1]{ {\rm e}^{-{\rm i} #1}}
\newcommand{\cre}[1]{\hat{#1}^\dagger}  %Creation operator
\newcommand{\des}[1]{\hat{#1}}  %Destruction operator

\newcommand{\op}[1]{\hat{#1}} %operator

\newcommand{\eqref}[1]{(\ref{#1})} %operator

\begin{document}
\title[Robust entangled states and MBQC using optical superlattices]{Creation of resilient entangled states and a resource for measurement-based quantum computation with optical superlattices}

\author{B.~Vaucher, A.~Nunnenkamp, and D.~Jaksch}

\address{Clarendon Laboratory, University of Oxford, Parks Road, Oxford OX1 3PU,
United Kingdom}

\ead{benoit.vaucher@merton.ox.ac.uk}
\submitto{\NJP}

\begin{abstract}
We investigate how to create entangled states of ultracold atoms trapped in optical lattices by dynamically manipulating the shape of the lattice potential. We consider an additional potential (the superlattice) that allows both the splitting of each site into a double well potential, and the control of the height of potential barrier between sites. We use superlattice manipulations to perform entangling operations between neighbouring qubits encoded on the Zeeman levels of the atoms without having to perform transfers between the different vibrational states of the atoms. We show how to use superlattices to engineer many-body entangled states resilient to collective dephasing noise. Also, we present a method to realize a 2D resource for measurement-based quantum computing via Bell-pair measurements. We analyze measurement networks that allow the execution of quantum algorithms while maintaining the resilience properties of the system throughout the computation.
\end{abstract}

\maketitle

\section{Introduction}

The experimental realization of bosonic spinor condensates in optical lattices has opened up the possibility to exploit the spin degree of freedom for a wide range of applications \cite{Gerbier2006a, Widera2006}. Recently, further experiments have proved that these systems are promising candidates for quantum information processing purposes (see~\cite{Anderlini2007} and references therein) and the study of quantum magnetism~\cite{Folling2007}. In the literature there are several proposals \col{using spinor condensates in optical lattice}, e.g.~to create macroscopic entangled states \cite{Pu2000, Duan2000} or to explore magnetic quantum phases~\cite{Rey2007}. In this paper, we consider the possibility of using optical superlattices to manipulate the spin degrees of freedom of the atoms and engineer many-body entangled states.

Superlattice setups allow the transformation of every site of the optical lattice into a double-well potential, and also the control of the potential barrier between sites~\cite{Peil2003,Lee2007,Folling2007}. Atoms with overlapping motional wave functions interact via cold collisions that coherently modify their spin, while conserving the total magnetization of the system~\cite{Ho1998, Widera2005}. Thus, by splitting each lattice site into a double-well potential, two atoms occupying the same site become separated by a potential barrier, and the interactions between them are switched off. Similarly, interactions between atoms can be switched on again for a certain time by lowering the potential barrier between the two sides of the double-well. Therefore, superlattice manipulations offer the possibility to apply operations between large numbers of atoms in parallel. In this work, we propose to exploit this parallelism to engineer highly-entangled many-body states.

The process of splitting a Bose-Einstein condensate using double-well potentials has already been studied in the context of mean-field theory~\cite{Milburn1997, Menotti2001, Jaksch2001}. In the first part of this paper, we will derive a two-mode effective Hamiltonian that allows an accurate description of the system as the superlattice is progressively turned on. Our approach differs from previous treatments of double well potentials (see e.g. Ref.~\cite{Milburn1997}) or the work in Ref.~\cite{Rey2007}, as the effective model we use is valid even for low barrier heights. We will use this effective Hamiltonian to analyze the process of splitting every site into a double-well potential starting with two atoms per site. We will show that this process creates a Bell state encoded on the Zeeman levels of the atoms in every site, each atom occupying one side of the double-well. Since these states are resilient to collective dephasing noise, we will use a lattice with a Bell pair in every site as a starting point to engineer many-body entangled states.

In the second part, we present a new method for implementing an entangling $\sqrt{\textrm{SWAP}}$ gate that does not require to transfer the atoms between different vibrational states. We will show that the application of this entangling gate between (or inside) Bell pairs allows the creation of many-body entangled states resilient to collective dephasing noise.

In the final part, we propose a new method involving superlattice manipulations \col{to realize} a 2D resource state for measurement-based quantum computating (MBQC) formally similar to a Bell-encoded cluster state. Since the realization of logical gates between non-neighbouring qubits is practically very difficult in optical lattices~\cite{Kay2004}, the one-way model for quantum computation is particularly relevant for these systems~\cite{Briegel2001,Raussendorf2001,Raussendorf2002}. The resource state we propose is resilient to collective dephasing noise, which makes it less prone to decoherence than the usual cluster states. Its utilization as a resource for MBQC requires adjustments of the measurement patterns used for cluster states, which we describe in the last section. We note that the realization of Bell-encoded cluster states as a resource for MBQC was proposed in Ref.~\cite{Tame2007} using lattice manipulations in three dimensions. The distinguishing feature of our proposal is that the resource state is created via 2D superlattice operations which have been demonstrated in the lab already.

This paper is organised as follows. In Sec.~\ref{sec:model} we derive an effective model for describing the dynamics of atoms within one site in the presence of a superlattice potential. In Sec.~\ref{sec:pairs}, starting with a lattice where every site contains two atoms of opposite spin, we show how to create a Bell pair in each site by splitting it using a superlattice potential. In Sec.~\ref{sec:gate}, we present how to perform a gate between two neighbouring qubits by manipulating the superlattice potential, and in Sec.~\ref{entangKnitting} and Sec.~\ref{sec:maxpersist} we show how to use this gate to create and probe many-body entangled states. Finally, in Sec.~\ref{sec:mbqc} we propose a method for creating a resource for MBQC via Bell-pair measurements and provide measurements network to implement a universal set of gates. We conclude in Sec.~\ref{sec:summary}.

%----------------------------------------------------------------------------------------

\section{Model}
\label{sec:model}
\image{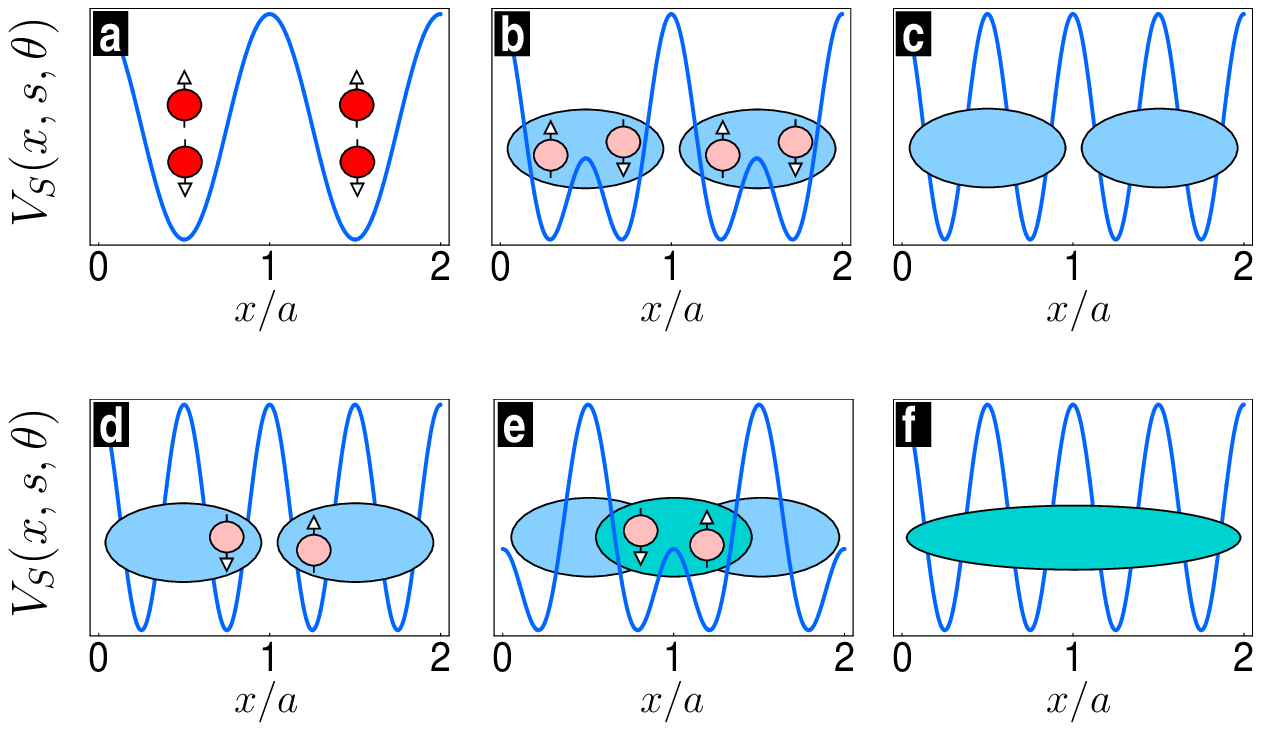}{12 cm}{(a) The system is initialized with two atoms per site in the state $w_a(\mathbf r)^2\otimes\mathcal S \left(\ket{f_1=1,m_1=+1;f_2=1,m_2=-1}\right)$ and the lattice parameters are $s=0$ and $\theta=0$. (b) The lattice potential is smoothly altered by increasing the lattice parameter $s$ while $\theta=0$. (c) At the end of the first step, the lattice parameter is $s=1$. Each lattice site then contains a Bell pair. (d-f) After setting the angle to $\theta=\pi/2$, the other set of barriers are lowered by decreasing the lattice parameter $s$ from $s=1$ (d) to $0 < s < 1$ (e) and back to $s=1$ (f).}{schematic}

We consider a gas of ultracold alkali atoms loaded into a deep optical lattice. The atoms are assumed to be bosons initialized in the $f=1$ hyperfine manifold~\cite{Widera2006}. The lattice potential is given by $V_{OL}(\mathbf r, s,\theta)=V_T[\sin^2(\pi z/a)+\sin^2(\pi y/a)+V_S(x,s,\theta)]$ where $V_T$ is the depth of the potential and 
\begin{equation}
V_S(x,s,\theta)=(1-s)\,\sin^2\left( \frac{\pi x}{a}+\theta \right) +s\, \sin^2\left( \frac{2 \pi x}{a}\right).
\end{equation}

%Each {\it site} of the lattice corresponds to a unit cell of size $a$ for $0 \le s <1$.
The depth of the potential will be given in units of the recoil energy $E_R=\hbar^2 \pi^2/(2M a^2)$ where $M$ is the mass of the atoms. The parameter $s\in [0,1]$ is determined by the relative intensities of the two pairs of lasers. The constant $a$ corresponds to the size of a unit cell when $s=0$ (see Fig.~\ref{schematic}a). We will refer to unit cells when $s=0$ as to {\it lattice sites}. Changing the value of $s$ from $0$ to $1$ transforms each site in a double well potential. We will refer to each side of this double well potential when $s=1$ as to a {\it subsite}. The angle $\theta$ allows for the manipulation of the potential barrier {\it inside} each site ($\theta=0$) or {\it between} sites ($\theta=\pi/2$).  A few lattice profiles corresponding to different parameters $s$ and $\theta$ are shown in Fig.~\ref{schematic}. In the following, we will refer to the lattice profiles corresponding to the values of $s\approx 0$ and $s\approx 1$ as to the {\it large} and {\it small} lattice limits, respectively.

We assume the lattice depth $V_T$ to be sufficiently large so that hopping can be neglected in the small and in the large lattice limits~\cite{Rey2004}.

%We assume that the hopping between sites and subsites can be neglected in both limits. This implies that the lattice has to be very deep, since the tunneling time in a lattice of depth $V_T$ with period $a/2$ is roughly $\sim 4 \exp(\sqrt{V_T/ E_R})$ times shorter than in a lattice with period $a$~\cite{Rey2004}.

The Hamiltonian of the system is given by
\begin{equation}\op{H}=\op{H}_K +\op{H}_Z+\op{H}_{\rm int}\label{htot}
\end{equation}
where $\op{H}_K$ describes the kinetic energy, $\op{H}_Z$ is the Zeeman term, and $\op{H}_{\rm int}$ describes the interactions between particles. These terms are defined by~\cite{Ueda2002, Zhang2004}
\begin{eqnarray}
\op{H}_K&=\int {\rm d}{\mathbf r} \sum_{\sigma=-1,0,1} \op{\Psi}_\sigma^\dag({\mathbf r})\, \op{h}_0(\mathbf r,s,\theta)\,\op{\Psi}_\sigma({\mathbf r}),\\
\op{H}_Z&=\int {\rm d}{\mathbf r} \sum_{\sigma=-1,0,1}  \Delta E_{Z,\sigma}(B)\,\op{\Psi}_\sigma^\dag({\mathbf r}) \op{\Psi}_\sigma({\mathbf r}),\\
\op{H}_{\rm int}&= \frac{1}{2}\int {\rm d}{\mathbf r} \Bigl[ c_0 \op{A}^\dag_{00}(\mathbf r) \op{A}_{00}(\mathbf r) + c_2 \sum_{m=-2}^2  \op{A}^\dag_{2m}(\mathbf r) \op{A}_{2m}(\mathbf r)\Bigr] ,\label{intHamilt}
\end{eqnarray}
where $c_0=4 \pi \hbar^2(2 a_2+ a_0 )/(3M)$ and $c_2=4 \pi \hbar^2(a_2-a_0 )/(3M)$ with $a_F$ the $s$-wave scattering length of the channel associated with the total angular momentum $F$, $\op{h}_0= (-\hbar^2 \mathbf \nabla^2/2M)+V_{OL}(\mathbf r,s,\theta)$ and $\Delta E_{Z,\sigma}(B)$ is the energy shift caused by the Zeeman effect in the presence of a magnetic field $\mathbf B=(0,0,B)$ oriented in the $z$--direction. The operator $\op{A}_{F m_F}(\mathbf r)$ is given by
\begin{equation}
\op{A}_{F m_F}(\mathbf r)=\sum_{m_1,m_2=-f}^f \braket{F,m_F}{f_1,m_1;f_2,m_2} \op{\Psi}_{m_1}(\mathbf r)\op{\Psi}_{m_2}(\mathbf r),
\end{equation}
where $\ket{f_1,m_1;f_2,m_2}=\ket{f_1,m_1}\otimes\ket{f_2,m_2}$ and $\braket{F,m_F}{f_1,m_1;f_2,m_2}$ is a Clebsch-Gordan coefficient.

In \ref{fullH}, we derive the effective Hamiltonian describing the dynamics of the atoms at one site using the field operator
\begin{equation}
\op{\Psi}^\dag_\sigma(\mathbf r) = \cre{a}_{\sigma} w_{a}(\mathbf r) +  \cre{b}_{\sigma} w_{b}(\mathbf r),
\label{fieldop}
\end{equation}
where $\cre{a}_\sigma$($\cre{b}_\sigma$) is the operator that creates a particle with spin $\ket{f=1,m_f=\sigma}$ ($\sigma=\{-1,0,1\}$ denotes the Zeeman level) in a motional state associated with the symmetric (anti-symmetric) mode function $w_{a}$($w_{b}$). The mode functions are centered in the middle of the site, and their shape depends on the lattice parameter $s$. The field operators obey the canonical bosonic commutation relations $[\op{\Psi}_\sigma({\mathbf r}),\op{\Psi}_{\sigma'}^\dag({\mathbf r'})] = \delta_{\sigma\sigma'}\delta(\mathbf{ r-r'})$ and $[\op{\Psi}_\sigma({\mathbf r}), \op{\Psi}_{\sigma'} ({\mathbf r'})] = [\op{\Psi}_\sigma^\dag ({\mathbf r}), \op{\Psi}_{\sigma'}^\dag({\mathbf r'})] = 0$. We will see in the next section that the inclusion of a second motional mode in the Hamiltonian allows an accurate description of the system's dynamics in both the large {\it and} the small lattice limit~\cite{Vaucher2007}. A more detailed explanation of the interaction term \eqref{intHamilt}, as well as the full Hamiltonian of the system in terms of creation (annihilation) operators can be found in \ref{fullH}.

%%--------------------------------------------------------------------------------------------------

\section{Creation of a Bell state on every lattice site}
\label{sec:pairs}

In this section, we use the effective Hamiltonian introduced in the previous section to show how to generate a lattice where every site contains a Bell state encoded on the Zeeman levels of the atoms. This procedure will be used as a preliminary step to engineer several types of many-body entangled states.

We start with a deep optical lattice in the large lattice limit (LLL) and two atoms per lattice site in the state
\begin{equation}
\ket{\psi_{\rm{init}}} = w_a(\mathbf r)^2\otimes\mathcal S \left(\ket{f_1=1,m_1=+1;f_2=1,m_2=-1}\right),\label{initS}
\end{equation}
where $\mathcal S$ is the symmetrization operator. In general, the two atoms can undergo spin-changing collisions coupling to various hyperfine levels. As we want to implement qubits on the Zeeman levels, \col{we will} limit the spin dynamics effectively to two hyperfine states.
There are several ways to achieve this in practice:
(i) One possibility is to exploit the conservation of angular momentum and choose two states such as $\ket{f=2, m_f=+2}$ and $\ket{f=1, m_f =+1}$, which are not coupled to any other hyperfine states by the interatomic interaction.
(ii) An accidental degeneracy in $^{87}\rm Rb$ offers yet another possible route: since the two $s$-wave scattering lengths $a_0$ and $a_2$ are almost equal, transitions between the state $\ket{f_1=1,m_1=0;f_2=1,m_2=0}$ and states where particles have opposite spins occur on a time scale $(c_0+c_2)/c_2 \approx 3 a_2/(a_2-a_0)\approx 300$ times slower than any other allowed transition. Hence, after initializing the system at time $t=0$ in state \eqref{initS}, we can limit our dynamical description to transitions between states with opposite spins, as long as the manipulations we are about to propose happen on a faster time scale than this transition. \col{Notice that the use of fast-switched microwave fields that suppress spin-changing collisions~\cite{Gerbier2006a} allows the relaxation of the constraint to be faster than the original time-scale of spin-changing collisions}. The preparation of the state \eqref{initS} has already been experimentally achieved \cite{Widera2005, Widera2006}.

The first step in our proposal consists of raising the superlattice potential, i.e.~changing the lattice parameter $s(t)$ from the LLL at time $t=0$ to the small lattice limit (SLL) at some time $t=T$ sufficiently slowly so that the ramp does not create excitations in the system. Once the superlattice is fully ramped up, each site is split in two, atoms no longer interact with each other and their spin state remains frozen in time.

The shape of the mode functions $w_a(\mathbf r)$ and $w_b(\mathbf r)$ depends on the lattice parameter. In the LLL, they correspond to the ground and first excited state of each lattice site. However, as the lattice parameter $s$ approaches the SLL, the mode functions transform into the symmetric and anti-symmetric superpositions $w_a(\mathbf r) = [\varphi_L(\mathbf r)+\varphi_R(\mathbf r)]/\sqrt{2}$ and $w_b(\mathbf r) = [\varphi_L(\mathbf r) - \varphi_R(\mathbf r)]/\sqrt{2}$, where the mode functions $\varphi_L(\mathbf r)$ and $\varphi_R(\mathbf r)$ are centered on either side of the double-well potential \cite{Vaucher2007,Milburn1997}. Hence, when the mode functions $w_{L,R}({\bf r})$ become centered within each subsite~\cite{Vaucher2007}, it is convenient to write the Hamiltonian of the system in terms of the operators 
\begin{eqnarray}
\cre{c}_{L,\sigma} &= (\cre{a}_{\sigma} + \cre{b}_{\sigma}) / \sqrt{2}\nonumber\\
\cre{c}_{R,\sigma} &= (\cre{a}_{\sigma} - \cre{b}_{\sigma}) / \sqrt{2},
\label{mtransfo}
\end{eqnarray}
which create a particle with spin $\sigma$ on the left and right side of the double-well potential, respectively. In the basis
\begin{eqnarray}
\ket{\uparrow \downarrow,} = \cre{c}_{L,\uparrow} \cre{c}_{L,\downarrow} \ket{\mathrm{vac}} & = & (\cre{a}_\uparrow + \cre{b}_\uparrow) (\cre{a}_\downarrow + \cre{b}_\downarrow)/2 \ket{\mathrm{vac}}, \nonumber\\
\ket{\uparrow,\downarrow} = \cre{c}_{L,\uparrow} \cre{c}_{R,\downarrow} \ket{\mathrm{vac}} & = & (\cre{a}_\uparrow + \cre{b}_\uparrow) (\cre{a}_\downarrow - \cre{b}_\downarrow)/2 \ket{\mathrm{vac}}, \nonumber \\
\ket{\downarrow, \uparrow} = \cre{c}_{L,\uparrow} \cre{c}_{R,\downarrow} \ket{\mathrm{vac}} & = & (\cre{a}_\uparrow - \cre{b}_\uparrow) (\cre{a}_\downarrow + \cre{b}_\downarrow)/2 \ket{\mathrm{vac}}, \nonumber \\
\ket{,\uparrow \downarrow} = \cre{c}_{R,\uparrow} \cre{c}_{R,\downarrow} \ket{\mathrm{vac}} & = & (\cre{a}_\uparrow - \cre{b}_\uparrow) (\cre{a}_\downarrow - \cre{b}_\downarrow)/2 \ket{\mathrm{vac}},
\end{eqnarray}
the Hamiltonian of the system reads (see \ref{fullH})
\begin{equation}
\op{H}_{\uparrow \downarrow} = \left( 
\begin{array}{cccc}
\tilde{E}+ \tilde{U}_{\uparrow \downarrow} & \tilde{J}_{\uparrow \downarrow} & \tilde {J}_{\uparrow \downarrow} & \chi_{\uparrow \downarrow}\\
\tilde{J}_{\uparrow \downarrow} & \tilde{E} + \chi_{\uparrow \downarrow} & \chi_{\uparrow \downarrow} & \tilde{J}_{\uparrow \downarrow}\\
\tilde{J}_{\uparrow \downarrow} & \chi_{\uparrow \downarrow} & \tilde{E} + \chi_{\uparrow \downarrow} & \tilde{J}_{\uparrow \downarrow}\\
\chi_{\uparrow \downarrow} & \tilde{J}_{\uparrow \downarrow} & \tilde {J}_{\uparrow \downarrow} & \tilde{E} + \tilde{U}_{\uparrow \downarrow}
\end{array}
\right),
\label{Ham_LR}
\end{equation}
where $\tilde{E} = V_a + V_b$ with $V_\nu= \int {\rm d}{\mathbf r}\, w^*_{\nu}(\mathbf r)\op{h}_0\,w_{\nu}(\mathbf r)$ and $\op{h}_0= (-\hbar^2 \mathbf \nabla^2/2M)+V_{OL}(\mathbf r,s,\theta)$ is the sum of the single-particle energies, $\tilde{U}_{\uparrow \downarrow} = \gamma_{\uparrow \downarrow} (U_{aa}+U_{bb}+6U_{ab})/4$ with $\gamma_{\uparrow \downarrow} U_{\nu\nu'}=\gamma_{\uparrow \downarrow} \int{\rm d}{\mathbf r} \,(|w_\nu(\mathbf r)| |w_{\nu'}(\mathbf r)|)^2$ and $\gamma_{\uparrow \downarrow}=c_0-c_2$ is the generalized on-site interaction, $\tilde{J}_{\uparrow \downarrow} = (V_a - V_b)/2 + \gamma_{\uparrow \downarrow} (U_{aa}-U_{bb})/4$ the generalized tunneling matrix element between the two sides of the well, and $\chi_{\uparrow \downarrow} = \gamma_{\uparrow \downarrow} (U_{aa}+U_{bb}-2U_{ab})/4$ corresponds to the density-density interaction energy between the two sides of the double-well. The latter is propor\-tional to the overlap between the functions $\varphi_{L}(\mathbf r)$ and $\varphi_{R}(\mathbf r)$. It cancels in the small lattice limit as a consequence of the localization of the mode functions $\varphi_{L}(\mathbf r)$ and $\varphi_{R}(\mathbf r)$ inside each subsite.

\image{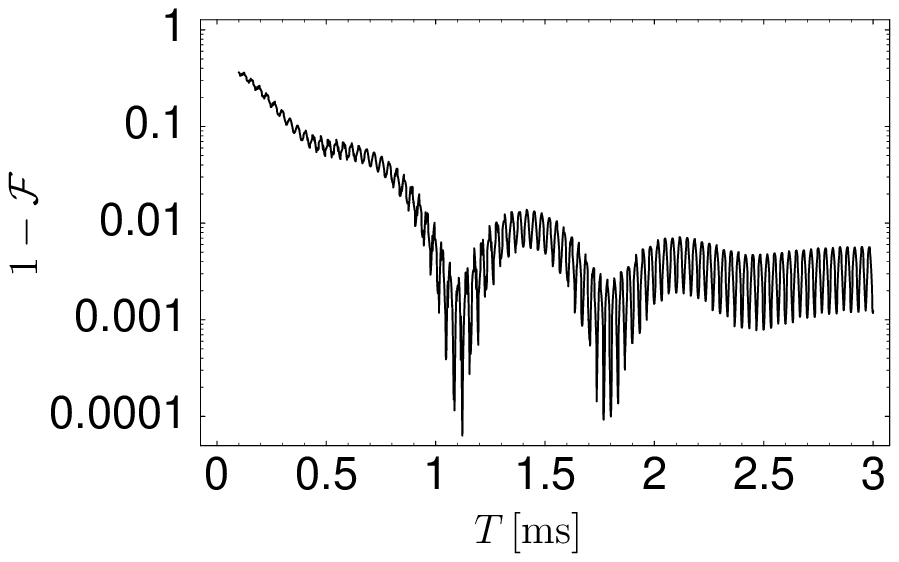}{8cm}{Numerical calculation of the anti-fidelity $1-\mathcal F$ where $\mathcal F = |\braket{\psi}{\psi_0}|^2$ with $\ket{\psi}$ the state of the two atoms at the end of the ramping. We have used the parameters of a lattice with period $a=\lambda/2$, $\lambda=840\,{\rm nm}$, and $V_T=60E_R$, initialized with two  $^{87}\rm Rb$ atoms per site in the state \eqref{initS} at time $T=0$ and a magnetic field of $B=30 \, {\rm G}$.}{fid}

Our two-mode model does not only describe the system well in both limits, it also takes into account the effects of density-density interactions between two subsites when the potential barrier between them is still not fully ramped up. This property makes it particularly valuable for the study of the system dynamics {\it between} the two limits. 

In the SLL, the Hamiltonian \eqref{Ham_LR} reduces (up to the energy shift $E$) to
\begin{equation}
\op{H}_{\uparrow \downarrow}^{\rm SLL}= \left( 
\begin{array}{cccc}
 U_{\uparrow \downarrow} & -J & -J & 0 \\
-J & 0 & 0 & -J \\
-J & 0 & 0 & -J \\
0 & -J & -J &  U_{\uparrow \downarrow}
\end{array}
\right).
\label{Ham_LR_2}
\end{equation}
where $V_a=E-J$ and $V_b=E+J$ with $E=\int {\rm d}{\mathbf r}\, \varphi^*_{L}(\mathbf r)\op{h}_0\,\varphi_{L}(\mathbf r)$ is the single-particle energy and $J=\int {\rm d}{\mathbf r}\, \varphi^*_{L}(\mathbf r)\op{h}_0\,\varphi_{R}(\mathbf r)$ the hopping integral, and $\tilde{U}_{\uparrow \downarrow}= \gamma_{\uparrow \downarrow} \int{\rm d}{\mathbf r} \,|\varphi_L|^4$ is the on-site interaction~\cite{Vaucher2007}.

For zero tunneling $J=0$, the Hamiltonian $\op{H}_{\uparrow \downarrow}^{\rm SLL}$ has two degenerate ground states: $\ket{\uparrow, \downarrow}$ and $\ket{\downarrow, \uparrow}$. This degenercy is lifted at finite $J$ where the symmetric superposition
\begin{equation}
\ket{\psi_0} = \frac{1}{\sqrt{2}}(\ket{\uparrow,\downarrow} +  \ket{\downarrow,\uparrow})
\label{miniGHZ}
\end{equation} 
is the ground and the antisymmetric superposition the first excited state. These two low-lying energy levels with an energy splitting of $4J^2/U_{\uparrow \downarrow}$ are separated from the other excited states by the interaction energy $U_{\uparrow \downarrow}$.

\col{We have carried out a dynamical simulation of the splitting dynamics using the exact {\it full} Hamiltonian defined in Eq. \eqref{htot} and \ref{fullH} for two particles and one site. For the time-scales considered throughout this paper, non-adiabatic effects due to the changes of shape of the mode functions can be safely neglected~\cite{Vaucher2007}}. We have used $s(t)=\sin^2[t/(2T)]$, and computed the anti-fidelity $1-\mathcal F$ with $\mathcal F= |\braket{\psi}{\psi_0}|^2$ between the state $\ket{\psi}$ of the system at the end of the numerical simulation of the ramp and the Bell state \eqref{miniGHZ}. For the parameters considered, we have found that the anti-fidelity reaches $1-\mathcal F\approx 10^{-3}$ for a quench time of $T\sim 1.1\,\pm 0.1\, \rm ms$ (see Fig.~\ref{fid}). It is expected that better fidelity can be obtained for larger values of $U_{\uparrow \downarrow}$~\cite{Teichmann2007}. This observation makes our scheme relevant for current experimental setups.

Thus, changing the topology of the optical lattice from the LLL to the SLL drives the state of the atoms within each lattice site from \eqref{initS} to the maximally entangled Bell state \eqref{miniGHZ}. Non-adiabatic transitions to excited states are suppressed by opposite parity in the case of the first excited state and by the on-site interaction $U_{\uparrow \downarrow}$ in the case of the other excited states. After this first step has been completed, a system of initially $N$ sites contains $k=2N$ subsites and its state is given by a tensor product of Bell pairs 
\begin{equation}
\ket{\Phi_k^{(0)}}=\bigotimes_{i=1}^N\, \ket{\psi_0}.\label{phiInit}
\end{equation}

Remarkably, the system in state \eqref{phiInit} is resilient to dephasing noise for a magnetic field slowly varying over a distance of one lattice site~\cite{Anderlini2007}.

%%----------------------------------------------------------------------

\section{Implementation of an entangling gate}
\label{sec:gate}

Since the interactions between two atoms depend on the overlap between their wave functions, they can be dynamically switched on and off by lowering and raising the potential barrier between the two subsites. This is done by varying the lattice parameter $s$ from $s=1$ at time $t=0$ to $s=1-\eta$ and back to $s=1$ at time $t=\tau$. In the SLL, where $\tilde {J}_{\uparrow \downarrow} /\tilde{U}_{\uparrow \downarrow} \ll 1$, tunneling can be treated perturbatively, and the Hamiltonian \eqref{Ham_LR} can be projected onto the subspace of singly-occupied sites.

We distinguish two cases: either the neighbouring subsites are occupied by atoms with opposite or equal spins. For two atoms with opposite spins, we can use Eq.~(\ref{Ham_LR}) and the effective Hamiltonian in the basis $\ket{\uparrow,\downarrow}$ and $\ket{\downarrow,\uparrow}$ reads (up to a constant energy shift) in second-order perturbation theory
\begin{equation}
\op{H}_{\uparrow \downarrow}^{(\rm eff)} = \left(\frac{2\tilde{J}_{\uparrow \downarrow} ^2}{\tilde{U}_{\uparrow \downarrow} } + \chi_{\uparrow \downarrow}  \right)  \left( \begin{array}{cccc} 1 & 1\\ 1 & 1 \end{array} \right).
\end{equation}
Since the effective Hamiltonian commutes with itself at different times, we obtain the following analytical expression for the time evolution operator
\begin{equation}
\op{U}(\tau) = \exp \left(-\frac{\rm i}{\hbar} \int_0^\tau H_{\uparrow \downarrow}^{(\rm eff)} \, {\rm d}t \right) = \left( \begin{array}{cccc} \expim{\phi} \cos \phi & -{\rm i} \expim{\phi} \sin \phi \\ -{\rm i} \expim{\phi} \sin \phi & \expim{\phi} \cos \phi \end{array} \right),
\end{equation}
where the phase $\phi$ is given by
\begin{equation}
\phi(\tau) = \frac{1} {\hbar} \int_0^\tau \left(\frac{2\tilde{J}_{\uparrow \downarrow}^2}{\tilde{U}_{\uparrow \downarrow}} + \chi_{\uparrow \downarrow} \right) \, {\rm d}t. \label{phitau}
\end{equation}
\col{While the first term in Eq.~\eqref{phitau} is due to second-order tunneling \cite{Rey2007, Folling2007}, the second term is due to density-density interactions. This term is negligible in the SLL, but when the potential barrier between the two sides of the double well is reduced, it adds an extra phase between neighbouring particles with different spins because of the difference between the scattering lengths $c_0$ and $c_2$.}

For atoms with equal spins $\sigma=\{\uparrow,\downarrow\}$ we start by writing down the basis states as

\begin{eqnarray}
\ket{\sigma \sigma,} = \cre{c}_{L,\sigma} \cre{c}_{L,\sigma} \ket{\mathrm{vac}} & = & (\cre{a}_\sigma + \cre{b}_\sigma) (\cre{a}_\sigma + \cre{b}_\sigma)/(2\sqrt{2}) \ket{\mathrm{vac}} \nonumber\\
\ket{\sigma,\sigma} = \cre{c}_{L,\sigma} \cre{c}_{R,\sigma} \ket{\mathrm{vac}} & = & (\cre{a}_\sigma + \cre{b}_\sigma) (\cre{a}_\sigma - \cre{b}_\sigma)/2 \ket{\mathrm{vac}} \nonumber \\
\ket{,\sigma \sigma} = \cre{c}_{R,\sigma} \cre{c}_{R,\sigma} \ket{\mathrm{vac}} & = & (\cre{a}_\sigma - \cre{b}_\sigma) (\cre{a}_\sigma - \cre{b}_\sigma)/(2\sqrt{2}) \ket{\mathrm{vac}},
\end{eqnarray}
in which the Hamiltonian reads
\begin{equation}
\op{H}_{\sigma \sigma} = \left( 
\begin{array}{ccc}
\tilde{E}_{\sigma \sigma} + \tilde{U}_{\sigma \sigma} & \tilde {J}_{\sigma \sigma} & \chi_{\sigma \sigma}\\
\tilde{J}_{\sigma \sigma} & \tilde{E}_{\sigma \sigma} + \chi_{\sigma \sigma} & \tilde{J}_{\sigma \sigma}\\
\chi_{\sigma \sigma} & \tilde {J}_{\sigma \sigma} & \tilde{E}_{\sigma \sigma} + \tilde{U}_{\sigma \sigma}
\end{array}
\right),
\end{equation}
where $\gamma_{\sigma \sigma} = c_0+c_2$ is the interaction constant for particles with equal spin $\sigma$, $\tilde{E}_{\sigma \sigma} = V_a + V_b \pm E_Z$ the sum of single-particle energies, $E_Z=- g\mu_B B\tau/2\hbar$ the Zeeman energy shift in the first order, $\tilde{J}_{\sigma \sigma} = (V_a - V_b)/\sqrt{2} + \gamma_{\sigma \sigma} (U_{aa}-U_{bb})/4$ the generalized tunneling matrix element, $\tilde{U}_{\sigma \sigma} = \gamma_{\sigma \sigma} (U_{aa}+U_{bb}+6U_{ab})/4$ the generalized on-site interaction and $\chi_{\sigma \sigma} = \gamma_{\sigma \sigma} (U_{aa}+U_{bb}-2U_{ab})/4$ the off-site density-density interaction.

In the SLL this reduces (up to the energy shift of $\tilde{E}_{\sigma \sigma}$) to
\begin{equation}
\op{H}_{\sigma \sigma}^{SLL} = \left( 
\begin{array}{cccc}
U_{\sigma \sigma}  & -\sqrt{2}J & 0 \\
-\sqrt{2}J & 0 & -\sqrt{2}J \\
0 & -\sqrt{2}J & U_{\sigma \sigma}
\end{array}
\right).
\end{equation}
For the state $\ket{\sigma, \sigma}$ there are no other low-energy states accessable, so that it acquires only the phase factor $\expi{\kappa}$ during the sweep. In second-order perturbation theory we find that $\kappa$ is similar to \eqref{phitau}
\begin{equation}
\kappa = \frac{1} {\hbar} \int_0^\tau \left( \frac{4\tilde{J}_{\sigma \sigma}^2}{\tilde{U}_{\sigma \sigma}} + \chi_{\sigma \sigma} \right) \, {\rm d}t. \label{kappatau}
\end{equation}
Since $\tilde{J}_{\uparrow \downarrow} \rightarrow -J$, $\tilde{J}_{\sigma \sigma} \rightarrow -\sqrt{2}J$ and $\gamma_{\uparrow \downarrow} /\gamma_{\sigma \sigma} = (c_0-c_2)/(c_0+c_2) \approx 1$, we get $\kappa \approx 2 \phi$.

Putting the results of this section together we find that lowering and raising the potential barrier between two subsites $n$ and $n+1$ implements the two-qubit gate $\op{U}$
\begin{eqnarray}
\fl
\op{U} \left(\begin{array}{c} 
\ket{00} \\ \ket{01} \\ \ket{10} \\ \ket{11} \\ 
\end{array} \right)
=
\left(\begin{array}{cccc}
\expi{E_Z \tau} & 0 & 0 & 0 \\
0 & \expi{\phi} \cos \phi & - \rm{i}\expi{\phi} \sin \phi & 0 \\
0 & - \rm{i}\expi{\phi} \sin \phi & \expi{\phi}\cos \phi & 0 \\
0 & 0 & 0 &  \expim{E_Z \tau}\\
\end{array} \right)
\left( \begin{array}{c} \ket{00} \\ \ket{01} \\ \ket{10} \\ \ket{11} \\ 
\end{array} \right).
\label{eqn:ugate}
\end{eqnarray}
where $\ket{ij}=\ket{i}_n \otimes \ket{j}_{n+1}$ with $\ket{1}_n =\cre{c}_{n,\uparrow}\ket{\rm vac}$ and $\ket{0}_n =\cre{c}_{n,\downarrow}\ket{\rm vac}$ where $\cre{c}_{n,\downarrow}$ and $\cre{c}_{n,\uparrow}$ are the creation operators of a particle at subsite $n$ with spin $\downarrow$ and $\uparrow$, respectively.

\col{Numerical simulations of the system dynamics have been carried out using the exact full Hamiltonian for two particles and one site}---that is, including the contribution of spin changing collisions to the $m_f=0$ state---for a system of $^{87}\rm Rb$ atoms, $V_T=60E_R$ and $\eta=0.3$. \col{We have computed the overlap between the states resulting from the numerical simulation with their analytical approximation} and found that for the parameters considered and switching times $\tau$ up to tens of milliseconds, the time evolution operator of the system is accurately approximated by Eq.~(\ref{eqn:ugate}). For longer times $\tau$, the evolution of the system in the reduced Hilbert becomes non-unitary due to slow transitions to the state where the two particles have spins $\sigma=0$, as expected.

Since the application of the gate $\op{U}$ is realized in parallel on all qubits in the $x$--direction, the phase due to the Zeeman shift in the first order in $B$ cancels for systems with an equal number of atoms in the $\sigma=+1$ and $\sigma=-1$ state.
%For these systems, only the parameter $\phi$ is of interest, and we will denote the operator $\op{U}$ by $\op{U}_\phi$.
For such systems, we find that for ramping times $\tau$ such that $\phi(\tau)=\{\pi/4,\pi/2, 3\pi/4\}$, lattice manipulations realize the gates (in the usual computational basis) \cite{Nielsen2000}
\begin{eqnarray}
\op{U}_{\frac{\pi}{4}}&=\sqrt{\textrm{SWAP}}\label{SqSwap}\\
\op{U}_{\frac{\pi}{2}}&=\textrm{SWAP},\label{SWAP}\\
\op{U}_{\frac{3\pi}{4}}&=\textrm{SWAP} \sqrt{\textrm{SWAP}}=\sqrt{\textrm{SWAP}}^\dag.\label{swapSqSwap}
\end{eqnarray} 

In the remainder of this paper we will show that, using a lattice of Bell pairs as a starting point, this gate\footnote{In the context of optical lattices, different methods to implement this gate have been put forward for different encodings of the logical qubits \cite{Hayes2007, Duan2003, Pachos2003}. Recently, a $\sqrt{\textrm{SWAP}}$ gate between qubits encoded on the internal spin state of the atoms was realized experimentally by manipulating both the spin and motional degrees of freedom of the atoms by means of optical superlattices~\cite{Anderlini2007}.} can be used to create both many-body entangled states and a resource state for MBQC.

%%--------------------------------------------------------------------------------------------------

\section{Creation and detection of a state with a tunable amount of entanglement}
\label{entangKnitting}

\image{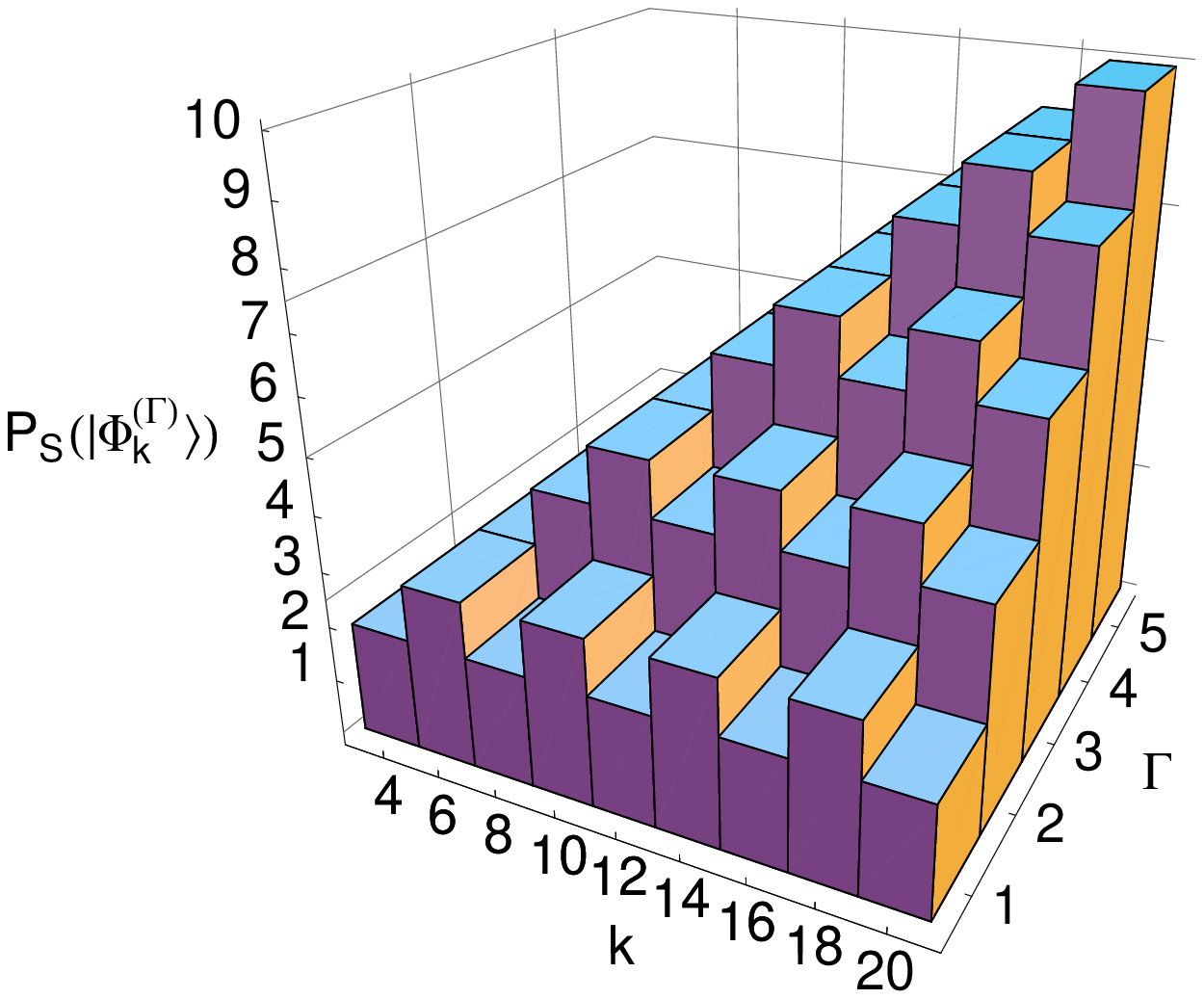}{8cm}{Schmidt measure of the state \eqref{globGamma} as a function of $\Gamma$ and $k$. The Schmidt decomposition of the state \eqref{globGamma} was calculated by partitioning the system in equal parts, which yields the maximum possible Schmidt rank.}{schmidtMes}

Depending on whether the lattice parameter $\theta$ is set to $\theta = 0$ or $\theta = \pi/2$, the gate $U_\phi$ in Eq.~\eqref{eqn:ugate} is applied between neighbouring subsites inside or between lattice sites, respectively. Up to an irrelevant global phase, these operations read
\begin{equation}
\op{\mathcal U}^{\rm in}(\op{U})=  \bigotimes_{n=1}^{N}\,\op{U} 
\label{opInside}
\end{equation}
or
\begin{equation}
\op{\mathcal U}^{\rm bw}(\op{U})=  \op{\mathbbm 1}\otimes \bigotimes_{n=1}^{N-1}\,\op{U} \otimes \op{\mathbbm 1}.
\label{opBetween}
\end{equation}
Using this notation, we define a {\it knitting} operator by
\begin{equation}
\op{\mathcal K}(\op{U})= \op{\mathcal U}^{\rm in}(\op{U})\op{\mathcal U}^{\rm bw}(\op{U}).
\label{globOp}
\end{equation}
Starting from a lattice of Bell pairs, the state resulting from $\Gamma$ successive applications of the operator $\op{\mathcal K}( \op{\mathcal U}_{\frac{\pi}{4}})$ on the state \eqref{phiInit} is denoted by
\begin{equation}
\ket{\Phi_k^{(\Gamma)}} = \op{\mathcal K}(\op{U}_{\frac{\pi}{4}})^\Gamma \ket{\Phi_k^{(0)}}.
\label{globGamma}
\end{equation}

In Fig.~\ref{schmidtMes}, we have plotted the Schmidt measure $P_{\rm s}^ L$ of the state \eqref{globGamma} for up to $k=20$ qubits and partitions of size $L=10$ as a function of $\Gamma$~\cite{Eisert2001,Briegel2001}. We find that the Schmidt rank of the state is directly proportional to $\Gamma$, i.e.~applying the entangling knitting operator (\ref{globOp}) first connects the initial Bell pairs and then further increases the entanglement content of the state. The largest possible Schmidt measure ${\rm max}[P_{\rm s}^ L(\ket{\Phi_k ^ {(\Gamma)}})] = k/2$ is reached after $\Gamma=N/2$ applications of the operator \eqref{globOp}. Once the maximum value for the Schmidt measure is reached, it remains unaffected by further applications of the operator \eqref{globOp}. We note that since the operator $\op{U}$ conserves the total number of spins, the state \eqref{globGamma} is resilient to collective dephasing noise. Recent experimental results suggest that resilience to this type of noise significantly increases the decoherence time of the system \cite{Anderlini2007}.

Since the application of the gate $\op{\mathcal K}(\op{U}_{\frac{\pi}{4}})$ affects the density correlations in the lattice, its effect on the system can be observed experimentally via state-selective measurement of the quasi-momentum distribution (QMD) of atoms with spin up~\cite{Greiner2002,Anderlini2007}
\begin{equation}
n_q^\uparrow=\frac{1}{2N}\sum_{i,j=1}^{2N} \, \expim{\pi q (i-j)/N}\langle \cre{c}_{i,\uparrow}\des{c}_{j,\uparrow}\rangle.\label{qdist}
\end{equation}
As an illustration, we have plotted in Fig.~\ref{cor} the QMD of a system of $k=14$ atoms after $\Gamma =1$ and  $\Gamma=7$ applications of the operator $\op{\mathcal K}(\op{U}_{\frac{\pi}{4}})$ on the state \eqref{phiInit}. Hence, starting from a lattice of Bell pairs, superlattice manipulations allow the creation of an entangled state with a tunable Schmidt rank that has a distinct experimental signature and is resilient to collective dephasing noise. These properties make this state suitable for experimental studies of many-body entanglement.

\image{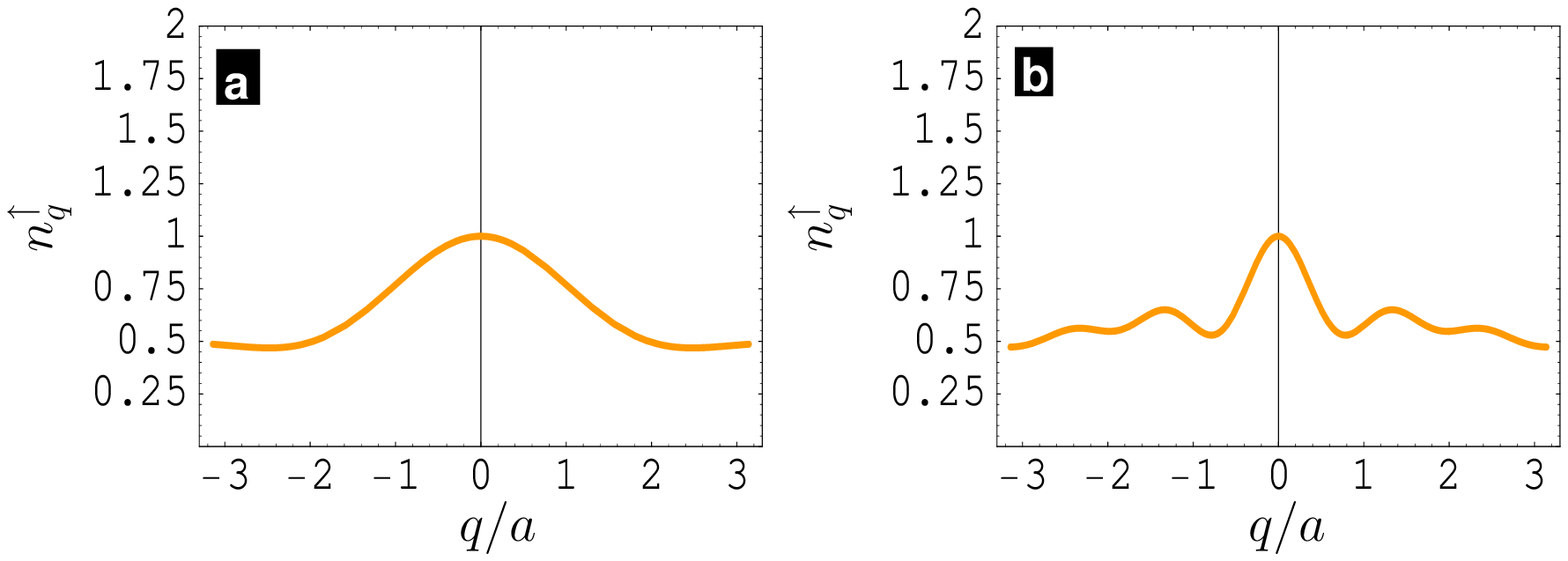}{12cm}{(a) Quasi-momentum distribution of $k=14$ atoms after the application of the gate $\op{\mathcal K}(\op{U}_{\frac{\pi}{4}})^\Gamma$ on the state \eqref{phiInit} for (a) $\Gamma=1$ and (b) $\Gamma=7$. After $\Gamma=7$ applications of the knitting operator, the state is maximally entangled.}{cor}

%%--------------------------------------------------------------------------------------------------

\section{Creation of maximally entangled states}
\label{sec:maxpersist}
Together with site-selective single-qubit operations applied in parallel on every second atom, superlattice manipulations can be used to implement the entangling phase-gate operation
\begin{equation}
\op{C}_{\rm i}= \op{U}_{\frac{\pi}{4}}(\op{Z}\otimes\op{\mathbbm 1})\op{U}_{\frac{\pi}{4}}(\op{\mathbbm 1}\otimes \op{Z}),
\label{PhaseGate}
\end{equation}
where $\op{C}_{\rm i}={\rm diag}(1,-\rm i,-\rm i,1)$ and $\op{Z}$ is the Pauli matrix defined as $\op{Z}={\rm diag}(1,-1)$ in the usual computational basis. The operation $\op{C}_{\rm i}$ is related to the operation $\op{C}_Z={\rm diag}(1,1,1,-1)$ by applying the single-qubit gate $\sqrt{\op{Z}}\otimes \sqrt{\op Z}$ on every site of the lattice. A proposal for realizing arbitrary single-qubit gates on individual atoms in an optical lattice can be found in Ref.~\cite{Zhang2006}. Notice that the single-qubit operations required to implement the gate \eqref{PhaseGate} between sites can be performed in parallel, and so each site need not be addressed separately. The implementation of a phase gate directly followed by a $\textrm{SWAP}$ gate
\begin{equation}
(\textrm{SWAP}) \, \op{C}_{i} =  \op{U}_{\frac{\pi}{4}}^\dag(\op Z\otimes\op{\mathbbm 1})\op{U}_{\frac{\pi}{4}}(\op{\mathbbm 1}\otimes \op Z) \label{swapPhase}
\end{equation}
is accomplished by adjusting the switching time $\tau$ of the last gate. The operation \eqref{swapPhase} is equivalent to $(\textrm{SWAP})\, \op{C}_Z$ up to the unitary transformation $\sqrt{\op Z}\otimes \sqrt{\op Z}$. Since the $\op{C}_Z$ operations between different qubits commute, $k/2$ successive applications of the operator $\op{\mathcal K}((\textrm{SWAP})\op{C}_Z)$ on the state $\ket{+}^{\otimes k}$ ($\ket{\pm}=(\ket{0}\pm \ket{1})/ \sqrt{2}$) create a complete graph state $\ket{K_N}$. Complete graphs have the property that each vertex is connected to all the vertices of the graph, i.e. a complete graph of $k$ vertices contains $k(k-1)/2$ edges. Graph states of complete graphs are equivalent to the maximally entangled state $\ket{GHZ}= (\ket{0}^{\otimes k} + \ket{1}^{\otimes k})/\sqrt{2}$ up to local unitary operations.

\image{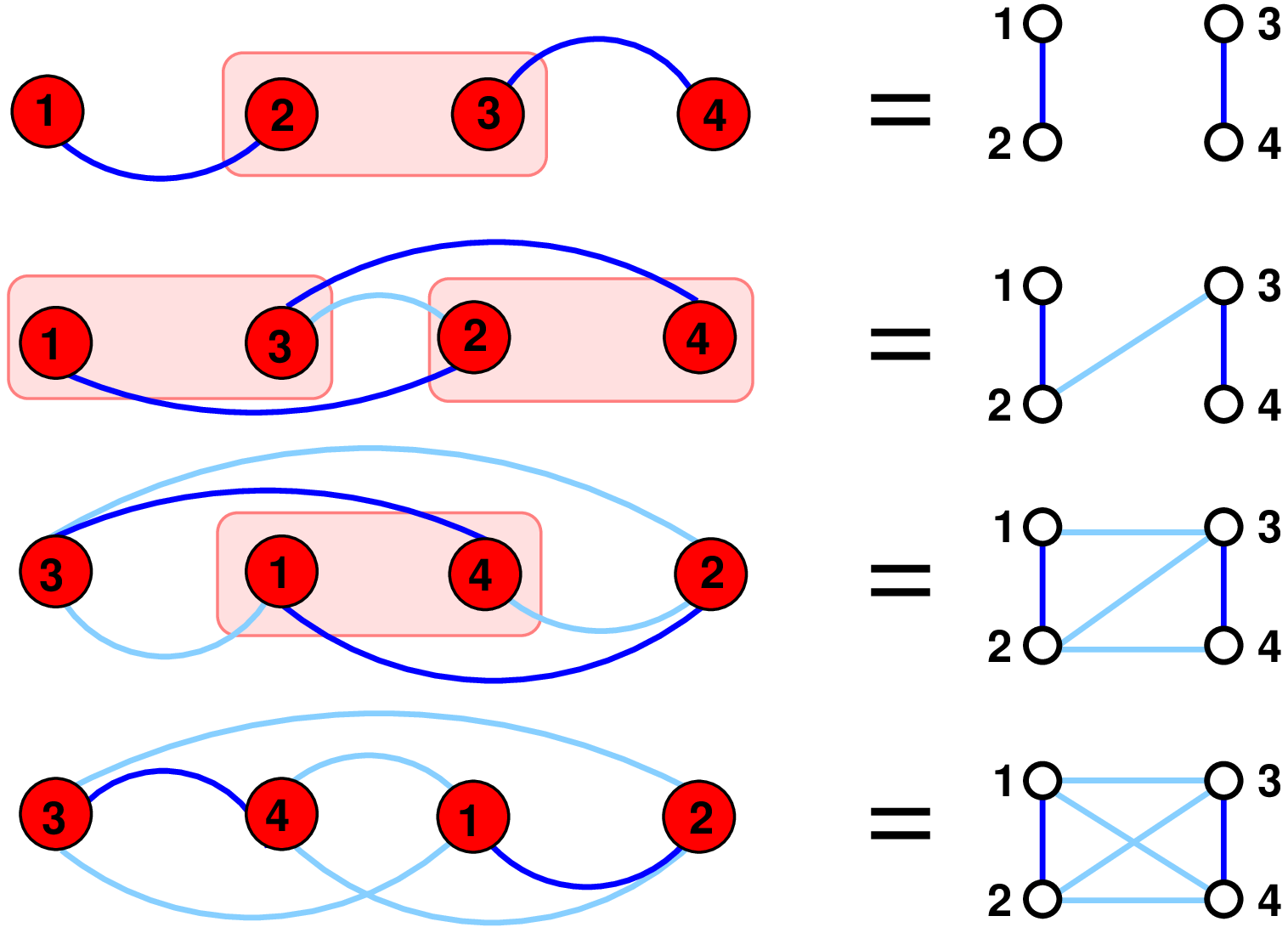}{8cm}{Schematic representations of the process leading to state \eqref{ourgraph}. Qubits initially forming a Bell pair are linked by a dark blue line. The boxes depict the location where the $\sqrt{SWAP}^\dag=SWAP \sqrt{SWAP}$ is performed. After a $\sqrt{SWAP}$ gate has been applied between two qubits, they are connected by a (light) blue line. Numerical calculations show that for up to 8 qubits, the state \eqref{ourgraph} is equivalent to a complete graph state under local unitary operations.}{completeGraph}

Superlattice manipulations alone allow the realization of the state
\begin{equation}
\ket{\Phi_k^{\rm graph}}=\op{\mathcal U}^{\rm bw}(\op{U}_{\frac{\pi}{4}}^\dag) \op{\mathcal K}(\op{U}_{\frac{\pi}{4}}^\dag)^{(k/2)-1} \ket{\Phi_k^{(0)}}
\label{ourgraph}
\end{equation}
resulting from $k-1$ successive applications of the operator~\eqref{swapSqSwap} between and inside each site. The process leading to state \eqref{ourgraph} is schematically represented in Fig.~\ref{completeGraph}. This state is resilient to collective dephasing noise, and possesses a structure similar to a complete graph state (see Fig.~\ref{completeGraph}). \col{Via the brute-force numerical calculation of the parameters of single-qubit unitary operations}, we have found that \eqref{ourgraph} is locally equivalent to a complete graph state of the same size for up to $k=\{4,6,8\}$ qubits. This suggests that this property holds for an arbitrary number of qubits.

The resilience of this state to collective dephasing noise, and its symmetry properties make it a good candidate for the improvement of the sensitivity of quantum spectroscopic measurements in noisy environments (see Ref.~\cite{Huelga1997}).

%%--------------------------------------------------------------------------------------------------

\section{Creation of a resource for MBQC}
\label{sec:mbqc}
In this section we will show how to create a state useful for MBQC via the application of the $\op{C}_Z$ gate via lattice manipulations in both the $x$ and $y$ direction. This state is formally similar to a Bell-encoded cluster state, and hence its utilization as a resource for MBQC only requires an adjustment of the measurement networks used for cluster states. To demonstrate the universality of our resource for MBQC and derive the measurement networks required to perform quantum algorithms, we will employ the method recently developed by Gross and Eisert in Ref.~\cite{Gross2007, Gross2007a}, which connects the matrix product representation (MPR) of a state with its computational power. A review of the basics concepts presented in Ref.~\cite{Gross2007} can be found in~\ref{appendix:gross}.

The MPR for a chain of $k$ systems of dimension $d=2$ (qubits) is given by
\begin{equation}
\ket{\psi_{\Lambda}}= \sum_{i_1 \dots i_k=\{0,1\}} \bra{R} \op{\Lambda}^{[i_1]}_1\cdots \op{\Lambda}^{[i_k]}_k\ket{L} \ket{i_1 \cdots i_k}.\label{MPS}
\end{equation}
It is specified by a set of $2k$ $D\times D$-matrices, which we will refer to as the correlation matrices, and two $D$-dimensional vectors $\ket{L}$ and $\ket{R}$ representing boundary conditions. The parameter $D$ is proportional to the amount of correlation between two consecutive blocks of the chain. Notice that the right boundary condition vector $\ket{R}$ appears on the left. This choice improves the clarity of calculations later when we will use the graphical notation explained in~\ref{appendix:grossGraphical}.

Starting from a lattice of Bell pairs, we find that the state resulting from the application of the gate $\op{\mathcal U}^{\rm bw}(\op{C}_Z)$ on the state $\ket{\Phi_k^{(0)}}$ \eqref{phiInit} has the MPS
\begin{equation}
\ket{\psi_{AB}} = \sum_{i_1 \dots i_k=\{0,1\}} \bra{R} \op{A}^{[i_1]} \op{B}^{[i_2]}\cdots \op{A}^{[i_{k-1}]} \op{B}^{[i_k]}\ket{L} \ket{i_1 \cdots i_k},
\label{ourMPS1}
\end{equation}
where
\begin{eqnarray}
\op{A}^{[0]}&=\ketbra{+}{0},\quad \op{A}^{[1]}&=\ketbra{-}{1},\label{Amat}\\
\op{B}^{[0]}&=\ketbra{1}{0},\quad \op{B}^{[1]}&=\ketbra{0}{1},\label{Bmat}\\
\ket{L}&=\ket{+},\quad \quad \ket{R}&=\sqrt{2}\ket{0}.
\end{eqnarray}
Here, each atom is labelled from $1$ to $k$, and the correlation matrices $\op{A}^{[0/1]}$ and $\op{B}^{[0/1]}$ are associated with odd and even atoms, respectively.
Equivalently, it can be written as
\begin{equation}
 \ket{\psi_C} = \sum_{\bar{i}_1 \dots \bar{i}_N=0}^{1} \bra{R} \op{C}^{[\bar{i}_1]}\cdots \op{C}^{[\bar{i}_N]}\ket{L} \ket{\bar{i}_1 \cdots \bar{i}_N},\label{ourMPS2}
\end{equation}
where
\begin{eqnarray}
\op{C}^{[\bar{0}]}&=\op{A}^{[1]}\op{B}^{[0]}=\ketbra{-}{0}\nonumber\\
\op{C}^{[\bar{1}]}&=\op{A}^{[0]}\op{B}^{[1]}=\ketbra{+}{1},\label{Cmat}\\
\ket{\bar{0}}&=\ket{10},\quad \ket{\bar{1}}=\ket{01}\nonumber.
\end{eqnarray}
In this representation, each site is labelled from $1$ to $N$ and the correlation matrix $\op{C}^{[\bar{0}/\bar{1}]}$ is associated with a pair of atoms.
The measurement of odd and even atoms in the $X$-eigenbasis $\mathcal{B}_{X}=\{\ket{\pm}=(\ket{0}\pm \ket{1})/ \sqrt{2} \}$ on the state \eqref{ourMPS1} projects the auxiliary matrix associated with the measured atom onto the eigenstate $(\ket{0}+ (-1)^s \ket{1})/ \sqrt{2}$ depending on the measurement outcome $s$ (see Eq.~\eqref{projMeasurement} in~\ref{appendix:gross}). In the graphical notation, this implements the operations
\begin{eqnarray}
\begin{xy}*!C\xybox{\xymatrix@C=5mm{\ar[r]&\xnode{A[X]}\ar[r]&}}\end{xy}& \propto \op{X}^s \op{H},\\
\begin{xy}*!C\xybox{\xymatrix@C=5mm{\ar[r]&\xnode{B[X]}\ar[r]&}}\end{xy}& \propto \op{X}\op{Z}^s,
\end{eqnarray}
for odd and even atoms, respectively. Here, the operator $\op{H}$ is the Hadamard gate and $\op{X}$ the Pauli-X matrix~\cite{Nielsen2000}. Similarly, measurements in the $\phi$-eigenbasis $\mathcal{B}_{\phi}=\{(\ket{0}\pm \expi{\phi}\ket{1})/ \sqrt{2} \}$ yield
\begin{eqnarray}
\begin{xy}*!C\xybox{\xymatrix@C=5mm{\ar[r]&\xnode{A[\phi]}\ar[r]&}}\end{xy}& \propto \op{X}^s \op{H} \op{S}(\phi) ,\\
\begin{xy}*!C\xybox{\xymatrix@C=5mm{\ar[r]&\xnode{B[\phi]}\ar[r]&}}\end{xy}& \propto \op{X}\op{Z}^s \op{S}(\phi),
\end{eqnarray}
where $\op{S}(\phi)={\rm diag}(1,\expi{\phi})$ in the usual computational basis. Also, we find that since the measurement of the $i$th and $(i+1)$th atoms ($i$ odd) in the Bell basis has only two possible outcomes, we have
\begin{eqnarray}
\begin{xy}*!C\xybox{\xymatrix@C=5mm{\ar[r]&\xnode{C[\bar{X}]}\ar[r]&}}\end{xy}& \propto \op{Z}\op{X}^s\op{H},\\
\begin{xy}*!C\xybox{\xymatrix@C=5mm{\ar[r]&\xnode{C[\bar{\phi}]}\ar[r]&}}\end{xy}& \propto \op{Z}\op{X}^s\op{H}\op{S}(\phi),
\end{eqnarray}
where the measurement eigenbases are given by $\mathcal{B}_{\bar{X}}=\{(\ket{\bar{0}}\pm \ket{\bar{1}})/ \sqrt{2} \}$ and $\mathcal{B}_{\bar{\phi}}=\{(\ket{\bar{0}}\pm \expi{\phi}\ket{\bar{1}})/ \sqrt{2} \}$, respectively. Since any single-qubit gate can be decomposed into Euler angles, i.e.
\begin{equation}
 \op{U}_{\rm rot}(\zeta,\eta,\xi)=\op{S}(\zeta)\op{H}\op{S}(\eta)\op{H}\op{S}(\xi),
\label{genSingleRot}
\end{equation}
the application of one of the following sequence of measurements
\begin{eqnarray}
\begin{xy}*!C\xybox{\xymatrix@C=5mm{\ar[r]&\xnode{\op{B}[X_1]}\ar[r]&\xnode{A[\phi_2]}\ar[r]&\xnode{B[\phi_3]}\ar[r]&\xnode{A[X_4]}\ar[r]&\xnode{B[\phi_5]}\ar[r]&\xnode{A[X_6]}\ar[r]&  }}\end{xy},\label{singleQrotAB}\\ 
\nonumber\\
\begin{xy}*!C\xybox{\xymatrix@C=5mm{\ar[r]&\xnode{C[\bar{X}_1]}\ar[r]&\xnode{C[\bar{\phi}_2]}\ar[r]&\xnode{C[\bar{\phi}_3]}\ar[r]&\xnode{C[\bar{\phi}_4]}\ar[r]&  }}\end{xy}\label{singleQrotC}
\end{eqnarray}
on the state \eqref{ourMPS1} realize arbitrary qubit gates, up to a known by-product operator $\op{U}_\Sigma$. For instance, applying the measurement sequence \eqref{singleQrotAB} with measurement outcomes $\vec{s}=(1,0,0,0,1,1)$, where $s_i$ denotes the outcome of $X_i$ or $\phi_i$, implements the operation $\op{U}_\Sigma \op{U}_{\rm rot} (\phi_2, \phi_3, \phi_5)$ with $\op{U}_\Sigma=\op{H}\op{Z}\op{X}$ on the state of the correlation space. Similarly, the measurement sequence \eqref{singleQrotC} with outcomes $\vec{s}=(0,0,1,0)$ implements the operation $\op{U}_\Sigma \op{U}_{\rm rot}(\phi_2,\phi_3,\phi_4)$ with $\op{U}_\Sigma=\op{Z}\op{X}\op{Z}$. Although both measurement sequences \eqref{singleQrotAB} and \eqref{singleQrotC} preserve the system's immunity to collective dephasing, less measurements are required using the sequence~\eqref{singleQrotAB}. The state of the correlation space can be read-out using a scheme described in~\ref{appendix:gross}.

\image{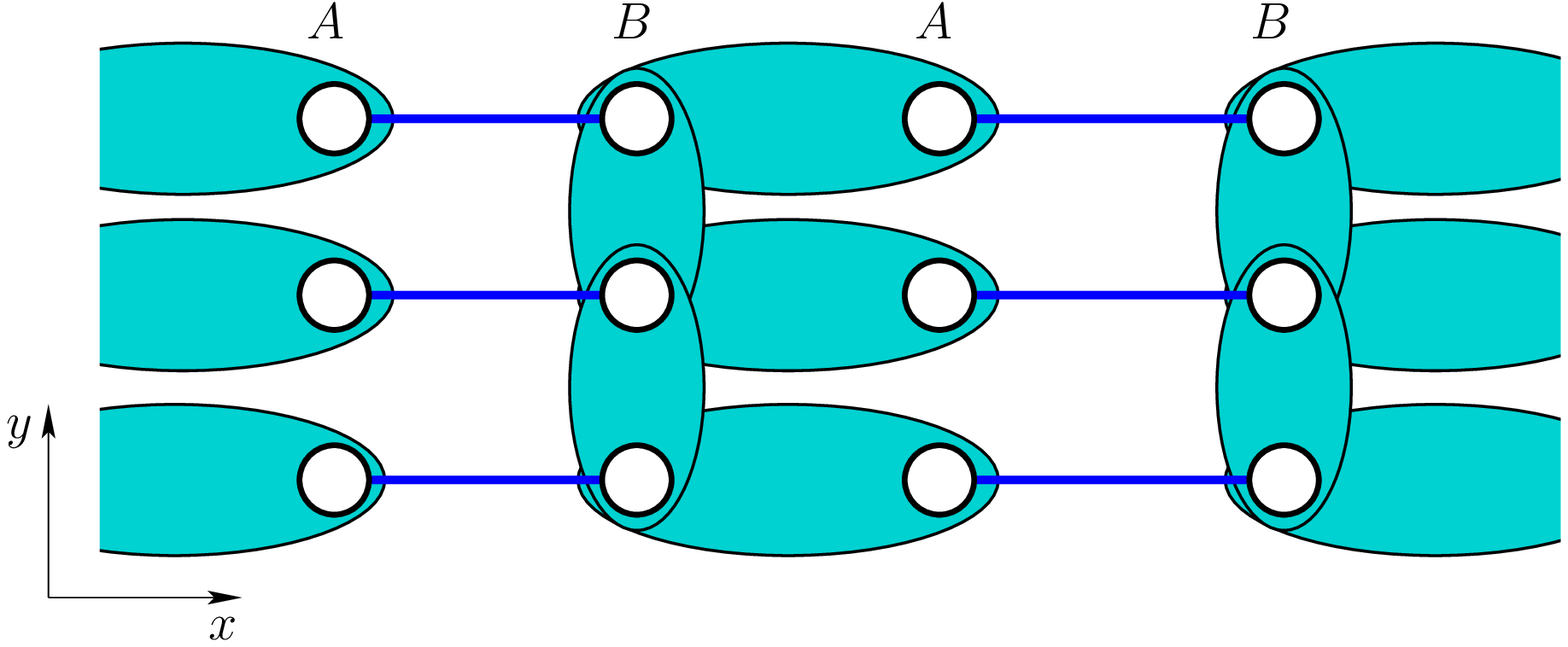}{10cm}{Starting with 1D system ($x$--direction) in the state \eqref{ourMPS1}, a 2D state useful for MBQC is obtained by applying $\op{C}_Z$ gates between even subsites (denoted by the letter $B$) in the $y$--direction.}{2dSetup}

Superlattice manipulations can also be used to engineer 2D systems. In the remainder of this section, we will consider the MPR of the 2D state $\ket{\psi_{2D}}$ resulting from the coupling of 1D systems in the state \eqref{ourMPS1} via the application of a $\op{C}_Z$ gate between {\it even} subsites\footnote{See e.g. Refs.~\cite{Sebby-Strabley2006,Anderlini2007} for relevant 2D superlattice setups} in the $y$--direction (see Fig.~\ref{2dSetup}). The MPR of this state is given by 
\begin{eqnarray}
 \begin{xy}*!C\xybox{\xymatrix@C=3mm@R=3mm{ && \\ \ar[r]^<<r&\xnode{C^{[0]}} \ar[r]^>>l \ar[u]^>>u& \\ & \ar[u]^<<d&}}\end{xy} & = \ket{-}_r \bra{0}_l \otimes \ket{+}_u\bra{0}_d,\label{C2D0}\\
 \begin{xy}*!C\xybox{\xymatrix@C=3mm@R=3mm{ && \\ \ar[r]^<<r&\xnode{C^{[1]}} \ar[r]^>>l \ar[u]^>>u& \\ & \ar[u]^<<d&}}\end{xy} & = \ket{+}_r\bra{1}_l\otimes \ket{-}_u\bra{1}_d.\label{C2D1}
\end{eqnarray}
Also, the expansion coefficients of $\ket{\psi_{2D}}$ are represented by
\begin{eqnarray}
\braket{\bar{i}_{1,1}\dots \bar{i}_{2,2}}{\psi_{2D}}&=& \begin{xy}*!C\xybox{
\xymatrix@C=3mm@R=3mm{ &\xnode{U} &\xnode{U} & \\ 
\xnode{L}\ar[r]&\xnode{C^{[\bar{i}_{1,1}]}}\ar[u]\ar[r]&\xnode{C^{[\bar{i}_{1,2}]}}\ar[r]\ar[u]&\xnode{R} \\
\xnode{L}\ar[r]&\xnode{C^{[\bar{i}_{2,1}]}}\ar[u]\ar[r]&\xnode{C^{[\bar{i}_{2,2}]}}\ar[r]\ar[u]&\xnode{R} \\
 &\xnode{D}\ar[u]&\xnode{D}\ar[u]&}
}\end{xy}
\end{eqnarray}
where $\ket{U}=\ket{0}$ and $\ket{D}=\ket{+}$.

This state is formally similar to a cluster state where qubits are encoded on two atoms and the $\ket{+}$ state corresponds to the Bell state~\eqref{miniGHZ}. Although the application of the $\op{C}_Z$ gate between {\it sites} in the $y$--direction implements the same operation between {\it encoded qubits}, it corresponds to a {\it different} operation between encoded qubits in the $x$--direction. Therefore, it is not possible to execute algorithms designed for cluster states using the state $\ket{\psi_{2D}}$ as a resource by simply interchanging, e.g. $X$ measurements with Bell-pair measurements $\bar{X}$. 

In the remainder of this section, we will show that $\ket{\psi_{2D}}$ constitute a universal resource for MBQC by presenting how to control information flows through the lattice, and by providing a measurement network that implements an entangling two-qubit gate.

Since the tensors of Eqs.~\eqref{C2D0} and \eqref{C2D1} factor, they can be represented graphically by
\begin{eqnarray}
 \begin{xy}*!C\xybox{\xymatrix@C=3mm@R=3mm{ && \\ \ar[r]^<<r&\xnode{C^{[0]}} \ar[r]^>>l \ar[u]^>>u& \\ & \ar[u]^<<d&}}\end{xy}&=&
\begin{xy}*!C\xybox{\xymatrix@C=3mm@R=3mm{
& &  & &\\
& &\xnode{+}\Pau^>>u& & \\
\Par^<<l&\xnode{0} &  &\xnode{-}\Par^>>r& \\
& &\xnode{0}& &\\
& & \Pau^<<d &  &}
}
\end{xy}.\label{factorC}
\end{eqnarray}
Using this representation and Eqs.~\eqref{C2D0} and \eqref{C2D1}, we find that
\begin{eqnarray}
\begin{xy}*!C\xybox{\xymatrix@C=3mm@R=3mm{
 &\xnode{C[\bar{Z}_1]}&  \\
\ar[r]&\xnode{C[\bar{X}_2]}\ar[u]\ar[r]&  \\
 &\xnode{C[\bar{Z}_3]}\ar[u]& }
}
\end{xy}&\propto& \op{X} \op{Z}^{s_1+s_2+s_3} \op{H}.
%\begin{xy}*!C\xybox{\xymatrix@C=3mm@R=3mm{ \ar[r]&\xnode{X Z^{s_1+s_2+s_3} H} \ar[r] &}}\end{xy}.\label{factorC}
\end{eqnarray}
Thus, measurements in the basis $\mathcal{B}_{\bar{X}}$ cause the information to flow from left to right, and measurements of the vertically adjacent sites in the $\mathcal{B}_{\bar{Z}}$ basis shields the information from the rest of the lattice~\cite{Gross2007}. Since we can isolate horizontal lines, they can be used as logical qubits.

An entangling two-qubit gate between two horizontal lines is realized as follows. Consider the following measurement network~\cite{Gross2007}
\begin{eqnarray}
\begin{xy}*!C\xybox{\xymatrix@C=3mm@R=3mm{
 &  & \\
\ar[r] &\xnode{C[\bar{X}_5]}\ar[r]&  \\
\xnode{C[\bar{Z}_3]}\ar[r]&\xnode{C[\bar{Y}_2]}\ar[u]\ar[r] &\xnode{C[\bar{Z}_4]} \\
\xnode{\ket{c}}\ar[r]&\xnode{C[\bar{X}_1]}\ar[u]\ar[r]& }
}
\end{xy}\,,\label{mNetwork}
\end{eqnarray}
where the middle site is measured in the basis $\mathcal{B}_{\bar{Y}}=\{(\ket{\bar{0}}\pm {\rm i} \ket{\bar{1}})/ \sqrt{2} \}$ and $c\in\{0,1\}$. The lower part of the network reduces to
\begin{eqnarray}
\begin{xy}*!C\xybox{\xymatrix@C=3mm@R=3mm{
 &  & \\
 \xnode{\ket{c}}\ar[r]&\xnode{C[\bar{X}_1]}\ar[u]\ar[r]& }
 }
 \end{xy}&\propto&
 \begin{xy}*!C\xybox{\xymatrix@C=3mm@R=3mm{
   &\\
 \xnode{Z^c \ket{+}}\ar[u]& & \\
 \xnode{(-1)^{s_1} X H \ket{c}}\ar[r]&}
}
\end{xy}.
\label{lowerPart}
\end{eqnarray}
Plugging \eqref{lowerPart} into the middle line yields
\begin{eqnarray}
\begin{xy}*!C\xybox{\xymatrix@C=3mm@R=3mm{
 &  & \\
\xnode{C[\bar{Z}_3]}\ar[r]&\xnode{C[\bar{Y}_2]}\ar[u]\ar[r] &\xnode{C[\bar{Z}_4]} \\
&\xnode{Z^c \ket{+}}\ar[u] & }
}
\end{xy}&\propto&
\begin{xy}*!C\xybox{\xymatrix@C=3mm@R=3mm{
   \\
\xnode{H Z^{s_2+s_3+c+1}[(-1)^{s_4}\ket{0}+{\rm i}\ket{1}]} \ar[u] }
}
\end{xy}.
\label{middlePart}
\end{eqnarray}
Finally, plugging \eqref{middlePart} into the upper of the network gives

\begin{eqnarray}
\begin{xy}*!C\xybox{\xymatrix@C=3mm@R=3mm{
\ar[r] &\xnode{C[\bar{X}_5]}\ar[r]&  \\
 & \xnode{H Z^{s_2+s_3+c+1}[(-1)^{s_4}\ket{0}+{\rm i}\ket{1}]} \ar[u]& }
}
\end{xy}&\propto& \op{U}_\Sigma (-{\rm i} \op{Z})^{c}
\label{upperPart}
\end{eqnarray}
where $\op{U}_\Sigma =\expi{\frac{\pi}{4}}({\rm i} \op{X})^{s_4} \op{H}\op{X} \op{S}(\frac{\pi}{2})^\dag \op{Z}^{s_5} (-{\rm i} \op{Z})^{s_2+s_3+1}$. Thus, the measurement network \eqref{mNetwork} implements an entangling controlled-phase gate between the upper and lower horizontal lines up to a know by-product operator. Since arbitrary single-qubit gates can be applied on each horizontal lines independently, this completes the proof of universality. The way of dealing with by-product operators at the end of the measurement sequences does not differ from the usual MBQC scheme with cluster states (see e.g.~\cite{Raussendorf2002}). Notice that since single and two-qubit gates are performed via the application of pairwise measurements, the system remains invariant to collective dephasing at any time during the execution of an algorithm.

%%--------------------------------------------------------------------------------------------------

\section{Summary}
\label{sec:summary}

We have analyzed the dynamics of a bosonic spinor condensate in a superlattice potential, and shown that a lattice with a Bell pair in every lattice site can be realized via the dynamical splitting of each site. We have proposed a scheme that allows the application of an entangling $\sqrt{\mathrm{SWAP}}$ gate between and inside Bell pairs. The successive application of this gate between and inside lattice sites was shown to create an entangled state with a tunable Schmidt rank that is resilient to collective dephasing noise; and a maximally entangled state which, as numerical evidence suggests, is locally equivalent to a GHZ state. Finally, we have presented a state that is obtained by connecting Bell pairs in two dimensions via an entangling phase gate, and shown that it constitutes a resource for MBQC formally similar to a Bell-encoded cluster state. We have provided measurement networks for implementing a two-qubit entangling gate as well as arbitrary local unitary operations. Our implementation has the advantage that it allows the execution of quantum algorithms while leaving the system unaffected by collective dephasing noise.

%%--------------------------------------------------------------------------------------------------

\ack
%\section{Acknowledgments}
This work was supported by the EU through the STREP project OLAQUI. The research was also supported by the EPSRC (UK) through the QIP IRC (GR/S82716/01) and EuroQUAM project EP/E041612/1. A.~N.~acknowledges a scholarship from the Rhodes Trust.

%********************************************************************
%                  APPENDIX
%********************************************************************

\appendix
\section*{Appendix}

\section{The full Hamiltonian}
\label{fullH}
Using the field operator \eqref{fieldop}, the single-particle terms of the Hamiltonian \eqref{htot} in second quantized form are given by
\begin{eqnarray}
%\fl %flush left
\op{H}_0 &= \op{H}_K + \op{H}_Z \nonumber\\
&= \sum_{\sigma=-1,0,1} [V_a +\Delta E_{Z,\sigma}(B)] \,\cre{a}_\sigma \des{a}_\sigma + \sum_{\sigma=-1,0,1} [V_b +\Delta E_{Z,\sigma}(B)] \,\cre{b}_\sigma \des{b}_\sigma,
\end{eqnarray}
where $V_\nu= \int {\rm d}{\mathbf r}\, w^*_{\nu}(\mathbf r)\op{h}_0\,w_{\nu}(\mathbf r)$. Assuming that the magnetic field is weak ($B< 200\,\rm G$), the Zeeman shift $\Delta E_{Z,\sigma}(B)$ is accurately approximated to the second order in $B$ \cite{Pethick2002}
\begin{equation}
\Delta E_{Z,\sigma}(B)=\cases{
\frac{g \mu_B B}{4}-\frac{3(g\mu_B B)^2}{16 \Delta E_{\rm hf}}& if $\sigma=-1$,\\
-\frac{(g\mu_B B)^2}{4 \Delta E_{\rm hf}}& if $\sigma=0$,\\
-\frac{g \mu_B B}{4}-\frac{3(g\mu_B B)^2}{16 \Delta E_{\rm hf}}& if $\sigma=1$,}
\end{equation}
where $g$ is the gyromagnetic factor, $ \Delta E_{\rm hf}$ is the hyperfine splitting energy, and $\mu_B$ is the Bohr magneton.

In the low-energy limit, atomic interactions are well approximated by $s$-wave scattering, with the scattering length depending on the spin state of the colliding atoms. At ultracold temperatures, two colliding atoms in the lower hyperfine level $f_{\rm low}$ will remain in the same multiplet, since the interaction process is not energetic enough to promote either atom to a higher hyperfine level $f_{\rm high}$~\cite{Ho1998}. Also, since alkalis have only two hyperfine multiplets, the conservation of the total angular momentum by the scattering process implies that interatomic interactions conserve the hyperfine spin $f_1$ and $f_2$ of the individual atoms. Thus, the interaction between two particles located at positions ${\mathbf r}_1$ and ${\mathbf r}_2$ is given by~\cite{Ho1998,Ueda2002,Pu1999}
\begin{equation}
\op{V}_{\rm int}({\mathbf r}_1-{\mathbf r}_2)=\frac{4 \pi \hbar^2 }{M }\delta({\mathbf r}_1-{\mathbf r}_2)\sum_{F=0,2}\, a_F \op{{\mathscr P}}_F,
\label{intOp}
\end{equation}
where $F=f_1+f_2$ is the total angular momentum of the scattered pair,  $\op{\mathscr P}_F$ is the projection operator for the total angular momentum $F$, and $a_F$ is the $s$-wave scattering length of the channel associated with the total angular momentum $F$. The operator $\op{\mathscr P}_F$ is given by~\cite{Pu1999}
\begin{equation}
\op{\mathscr P}_F=\sum_{m_F=-F}^F\,\ket{F,m_F}\bra{F,m_F},
\end{equation}
where $\ket{F,m_F}$ is the state formed by two atoms with total angular momentum $F$ and $m_F=m_1+m_2$ with $m_1$ and $m_2$ the projection on the quantization axis of $f_1$ and $f_2$, respectively. The quantization axis is defined by the direction of the magnetic field present in the system. Boson statistics require the state $\ket{F,m_F}$ to be invariant under particle exchange, and hence only the terms corresponding to values of $F=\{0,2\}$ appear in the sum of Eq.~\eqref{intOp}. Using the relation $\op{\mathbf f}_1 \cdot \op{\mathbf f}_2=\op{\mathscr P}_2-2 \op{\mathscr P}_0$, where $\op{\mathbf f}_i=(\op{f}_x^i,\op{f}_y^i,\op{f}_z^i)$ is the total angular momentum operator of the particle $i$, the interaction operator can be re-written as~\cite{Ho1998,Stamper-Kurn2001} 
\begin{equation}
\op{V}_{\rm int}({\mathbf r}_1-{\mathbf r}_2)= \delta({\mathbf r}_1-{\mathbf r}_2)\left( c_0+c_2 \op{\mathbf f}_1 \cdot \op{\mathbf f}_2 \right). \label{intSimple}
\end{equation}
From Eq.~\eqref{intSimple} we see that interactions conserve the quantum number $m_F$ \cite{Pethick2002, Bigelow2005}.

Using the field operator \eqref{fieldop}, the interaction term in second quantized form reads
\begin{eqnarray}
\op{H}_{\rm int}&= \frac{1}{2} \int \int {\rm d}{\mathbf r_1}{\rm d}{\mathbf r_2}\, \delta({\mathbf r}_1-{\mathbf r}_2) \times\nonumber\\
&c_0\,:\left( \op{\mathbf \Psi}^\dag(\mathbf r_1)\cdot \op{\mathbf \Psi}(\mathbf r_1)\right)\left( \op{\mathbf \Psi}^\dag(\mathbf r_2) \cdot \op{\mathbf \Psi}(\mathbf r_2)\right): \nonumber\\
&+  c_2  \sum_{\ell=x,y,z} :\left( \op{\mathbf \Psi}^\dag(\mathbf r_1) \op{f}_{\ell}^1 \op{\mathbf \Psi}(\mathbf r_1)\right)\left( \op{\mathbf \Psi}^\dag(\mathbf r_2) \op{f}_{\ell}^2 \op{\mathbf \Psi}(\mathbf r_2)\right): ,
\label{vecHint}
\end{eqnarray}
where $:\op{\circ}:$ represents the operator $\op{\circ}$ in normal ordering, $\op{\mathbf \Psi}(\mathbf r)=(\op{\Psi}_{-1}(\mathbf r),\op{\Psi}_0(\mathbf r),\op{\Psi}_{1}(\mathbf r))$ and the matrices $ \op{f}_{\ell}$ are given by~\cite{Zhang2004}
\begin{equation}
\fl
\op{f}_{x}^i = \frac{1}{\sqrt{2}}\left( \begin{array}{ccc} 0&1&0 \\ 1&0&1 \\ 0&1&0 \end{array} \right), \,
\op{f}_{y}^i = \frac{1}{\sqrt{2}\rm i}\left( \begin{array}{ccc} 0&1&0 \\ -1&0&1 \\0&-1&0 \end{array} \right), \,
\op{f}_{z}^i = \left( \begin{array}{ccc} 1&0&0 \\ 0&0&0\\0&0&-1\end{array}\right).
\end{equation}
The Hamiltonian given in \eqref{intHamilt} is a compact form of \eqref{vecHint}. In its expanded form, Eq.~\eqref{vecHint} reads~\cite{Stamper-Kurn2001}
\begin{eqnarray}
\op{H}_{\rm int}=& \frac{1}{2}\int {\rm d} {\mathbf r} \nonumber\\
& \Bigl\{ (c_0 + c_2) \left[  \op{\Psi}^\dag_1 (\mathbf r)  \op{\Psi}^\dag_1 (\mathbf r) \op{\Psi}_1 (\mathbf r)   \op{\Psi}_1 (\mathbf r)+ \op{\Psi}^\dag_{-1} (\mathbf r)  \op{\Psi}^\dag_{-1}(\mathbf r) \op{\Psi}_{-1}(\mathbf r)   \op{\Psi}_{-1} (\mathbf r) \right] \nonumber\\ 
& + 2(c_0 + c_2) \left[  \op{\Psi}^\dag_1 (\mathbf r)  \op{\Psi}^\dag_0 (\mathbf r) \op{\Psi}_1 (\mathbf r)   \op{\Psi}_0 (\mathbf r)+ \op{\Psi}^\dag_{-1} (\mathbf r)  \op{\Psi}^\dag_{0}(\mathbf r) \op{\Psi}_{-1}(\mathbf r)   \op{\Psi}_{0} (\mathbf r) \right] \nonumber\\
 & + 2 c_2 \left[  \op{\Psi}^\dag_0 (\mathbf r)  \op{\Psi}^\dag_0 (\mathbf r) \op{\Psi}_1 (\mathbf r)   \op{\Psi}_{-1} (\mathbf r)+ \op{\Psi}^\dag_1 (\mathbf r)  \op{\Psi}^\dag_{-1}(\mathbf r) \op{\Psi}_0(\mathbf r)   \op{\Psi}_0(\mathbf r) \right] \nonumber\\ 
& + c_0 \, \left[ \op{\Psi}^\dag_0 (\mathbf r)  \op{\Psi}^\dag_0 (\mathbf r) \op{\Psi}_0 (\mathbf r)   \op{\Psi}_0 (\mathbf r)\right] \nonumber\\
& + 2(c_0-c_2) \, \left[\op{\Psi}^\dag_1 (\mathbf r)  \op{\Psi}^\dag_{-1}(\mathbf r) \op{\Psi}_1 (\mathbf r)   \op{\Psi}_{-1}(\mathbf r)\right]\Bigr\}.\label{expandedHint}
\end{eqnarray}

The interaction Hamiltonian \eqref{expandedHint} can be expressed in terms of $\cre{a}_\sigma$ and $\cre{b}_\sigma$ operators using the relation
\begin{eqnarray}
\fl %flush left
\frac{1}{2}\int {\rm d}{\mathbf r}\, \op{\Psi}^\dag_{\sigma}(\mathbf r)  \op{\Psi}^\dag_{\gamma} (\mathbf r) \op{\Psi}_{\sigma'} (\mathbf r)   \op{\Psi}_{\gamma'} (\mathbf r) = \frac{U_{aa}}{2} \left(  \cre{a}_{\sigma}\cre{a}_{\gamma}\des{a}_{\sigma'}\des{a}_{\gamma'} \right)+\frac{U_{bb}}{2} \left(  \cre{b}_{\sigma}\cre{b}_{\gamma}\des{b}_{\sigma'}\des{b}_{\gamma'} \right)+\nonumber\\
\frac{U_{ab}}{2} \left(  \cre{a}_{\sigma}\cre{a}_{\gamma}\des{b}_{\sigma'}\des{b}_{\gamma'} + \cre{b}_{\sigma}\cre{b}_{\gamma}\des{a}_{\sigma'}\des{a}_{\gamma'} \right)
+ \frac{U_{ab}}{2}\left(  \cre{a}_{\sigma}\cre{b}_{\gamma}+   \cre{b}_{\sigma}\cre{a}_{\gamma}\right)\left(  \des{a}_{\sigma'}\des{b}_{\gamma'}+   \des{b}_{\sigma'}\des{a}_{\gamma'}\right),
\end{eqnarray}
where $U_{\nu\nu'}=\int{\rm d}{\mathbf r} \,(|w_\nu(\mathbf r)| |w_{\nu'}(\mathbf r)|)^2$ and $\sigma,\gamma=-1,0,+1$.

%-------------------------------------------------------------------------------
\section{Measurement-based quantum computations and MPSs}
\label{appendix:gross}
\subsection{General idea}
In this section, we present the general idea developed in Ref.~\cite{Gross2007a}. From Eq.~\eqref{MPS}, one can see that if the $i_k$ site in the computational basis and the outcome $s_k$ is obtained, then the state $\ket{L}$ of the auxiliary system becomes
\begin{equation}
 \ket{L}'=\op{\Lambda}^{[s_k]}_k \ket{L}.
\end{equation}
Thus, measurements change the state $\ket{L}$ of the auxiliary system. Similarly, if the measurement of a local observable at site $k$ yields an outcome corresponding to the observable's eigenvector $\ket{\phi_k}$, then the matrix $\op{\Lambda}^{[i_k]}_k$  transforms into
\begin{equation}
\op{\Lambda}[\phi]_k= \braket{\phi_k}{0}\op{\Lambda}^{[0]}_k + \braket{\phi_k}{1}\op{\Lambda}^{[1]}_k,\label{projMeasurement}
\end{equation}
and the vector of the auxiliary system becomes
\begin{equation}
\ket{L}' = \op{\Lambda}[\phi]_k \ket{L}.
\end{equation}
From this point of view, a measurement on some physical site change the correlation properties between this site and the rest of the chain, i.e.~it performs an operation on the auxiliary system state $\ket{L}$, which we will sometimes refer to as the state of the correlation space.
\subsection{Graphical notation}
\label{appendix:grossGraphical}
In order to deal with the MPS of higher-dimensional states, the graphical notation introduced in Ref.~\cite{Gross2007a} is very helpful. In this notation, tensors are represented by boxes, and their indices by edges. Vectors and matrices are thus symbolized by
\begin{eqnarray}
\ket{L}&=&\xymatrix@C=5mm{\xnode{L}\ar[r]&},\\
\bra{R}&=&\xymatrix@C=5mm{\ar[r]&\xnode{R^\dag}&},\\
\op{\Lambda}^{[i]}&=&\xymatrix@C=5mm{\ar[r]&\xnode{\Lambda^{[i]}}\ar[r]&}.
\end{eqnarray}
Consequently, vector and matrix operations can be represented graphically by
\begin{eqnarray}
\braket{R}{L}&=&\xymatrix@C=5mm{\xnode{L}\Par&\xnode{R}}\quad,\\
\op{B}\op{A}&=&\xymatrix@C=5mm{\ar[r]&\xnode{A}\ar[r]&\xnode{B}\ar[r]&},\\
\op{B}\op{A}\ket{L}&=&\xymatrix@C=5mm{\xnode{L}\ar[r]&\xnode{A}\ar[r]&\xnode{B}\ar[r]&}.
\end{eqnarray}
\subsection{Read-out scheme}
In order to use the state of the correlation space to process quantum information, we must be able to read it out at the end of the computation. It turns out that in our case a measurement of the $(i-1)$th site in the computational basis corresponds to a measurement of the correlation system just after the $i$th site. To illustrate this fact, assume that just after the measurement of the $i$th site, the correlation system is in the state $\ket{0}$, that is
\begin{eqnarray}
\begin{xy}*!C\xybox{\xymatrix@C=5mm{\xnode{L}\ar[r]&\xnode{B[\phi_1]}\ar[r]&\xnode{A[\phi_2]}\ar[r]&...&\xnode{B[\phi_i]}\ar[r]&}}\end{xy} & = \ket{0},
\end{eqnarray}
or
\begin{eqnarray}
\begin{xy}*!C\xybox{\xymatrix@C=5mm{\xnode{L}\ar[r]&\xnode{B[\phi_1]}\ar[r]&\xnode{A[\phi_2]}\ar[r]&...&\xnode{A[\phi_i]}\ar[r]&}}\end{xy} & = \ket{0}.
\end{eqnarray}
Thus, using Eqs.~\eqref{Amat} and \eqref{Bmat} we have
\begin{eqnarray}
 \begin{xy}*!C\xybox{\xymatrix@C=5mm{\xnode{\ket{0}}\ar[r]&\xnode{A[1]}\ar[r]&}}\end{xy} \propto \ket{-}\braket{1}{0} &=&0,\\
\begin{xy}*!C\xybox{\xymatrix@C=5mm{\xnode{\ket{0}}\ar[r]&\xnode{B[1]}\ar[r]&}}\end{xy} \propto \ket{0}\braket{1}{0} &=&0.
\end{eqnarray}
Hence, the probability of obtaining the result $1$ for a measurement on the $(i-1)$th site is zero. Consequently, if the state of the correlation system is in state $\ket{0}$ after the $i$th site, then the $(i-1)$th physical site must also be in that state. Since
\begin{eqnarray}
 \begin{xy}*!C\xybox{\xymatrix@C=5mm{\xnode{\ket{0}}\ar[r]&\xnode{C[1]}\ar[r]&}}\end{xy} & \propto \ket{+}\braket{1}{0}=0,
\end{eqnarray}
the same observation applies if one measures the $(i-2)$th and $(i-1)$th sites ($i$ odd) in the basis $\mathcal{B}_{\bar{Z}}=\{ \ket{\bar{0}},\ket{\bar{1}} \}$. Therefore, as a similar argument applies if the state of the correlation system is in the state $\ket{1}$ after the $i$th site, the description of the read-out scheme is complete.

%---------------------------------------------------------------------------------
\section*{References}
\bibliography{article}
%\bibliography{/home/vaucher/bibliography/main}
\end{document}